\documentclass[12pt]{iopart}

\usepackage{iopams}  
\usepackage[english]{babel}
\usepackage{graphicx}
\usepackage{cite}
\usepackage{bbm}
 
\begin{document}

%
\title{Operator approach to quantum optomechanics}

\author{C. Ventura-Vel\'azquez}
\address{Instituto Nacional de Astrof\'{\i}sica, \'Optica y Electr\'onica, Calle Luis Enrique Erro No. 1, Sta. Ma. Tonantzintla, Pue. CP 72840, M\'exico}

\author{B.~M. Rodr\'{\i}guez-Lara}
\address{Instituto Nacional de Astrof\'{\i}sica, \'Optica y Electr\'onica, Calle Luis Enrique Erro No. 1, Sta. Ma. Tonantzintla, Pue. CP 72840, M\'exico}
\ead{bmlara@inaoep.mx}

\author{H.~M. Moya-Cessa}
\address{Instituto Nacional de Astrof\'{\i}sica, \'Optica y Electr\'onica, Calle Luis Enrique Erro No. 1, Sta. Ma. Tonantzintla, Pue. CP 72840, M\'exico}

%
\begin{abstract}
We study the mirror-field interaction in several frameworks: when it is
driven, when it is affected by an environment and when a two-level atom is introduced
in the cavity. By using operator techniques we show how these problems may be either
solved or how the Hamiltonians involved, via sets of unitary transformations, may be
taken to known Hamiltonians for which there exist approximate solutions.
\end{abstract}


\maketitle

%
\section{Introduction}

Light carries momentum and, therefore, it can exert pressure over matter \cite{Poynting1909p560,Poynting1910}, be it from incoherent \cite{Lebedev1901p433,Nichols1901p307,Lebedev1910p411} or coherent \cite{Letokhov1968p348,Ashkin1970p156,Dalibard1984p4577} sources.
Such radiation pressure allows, for example, the coupling of mechanical degrees of freedom to electromagnetic cavity modes in cavities with a moving mirror, both in the classical \cite{Braginskii1967p986,Dorsel1983p1550,Gozzini1985p1841} and quantum regimes \cite{Braginskii1967p1434,Braginskii1969p1421,Caves1980p75,Jacobs1994p1961,Law1995p2537}.
The so-called standard optomechanical model in quantum optics is modeled after a classical Fabry-P\'erot resonator where one mirror is free to move in a pendulum like motion \cite{Braginskii1967p1434}.
In the beginning, the interest on this optomechanical system was focused on the detection of gravitational waves.
When the effects of radiation pressure on the device were shown to be a detection issue \cite{Caves1980p75}, it became important to beat the standard quantum limit \cite{Caves1981p1693,Pace1993p3173}.
An interesting solution to this problem is to prepare the mechanical oscillator in a non-Gaussian state \cite{Mancini1994p4055,Pace1993p3173,Khalili2010p070403}, thus quantum state engineering of the mechanical mode became important. 
Furthermore, it is of great interest to test quantum theory with macroscopic degrees of freedom \cite{Leggett1980p80,Bose1999p3204,Poot2012p273} and quantum optomechanical systems provide an experimentally feasible testing ground and may even be a viable quantum information platform \cite{Kippenberg2008p1172,Marquardt2009p40}.

The canonical quantization of a Fabry-P\'erot cavity with a pendulum-like mirror delivers an ideal Hamiltonian in the form \cite{Moore1970p2679,Caves1980p75,Meystre1985p1830,Law1995p2537}, 
\begin{eqnarray}
\hat{H} = \omega_{c}\hat{a}^{\dag}\hat{a}+\omega_{m}\hat{b}^{\dag}\hat{b}-g\hat{a}^{\dag}\hat{a}\left(\hat{b}^{\dag}+\hat{b}\right),
\end{eqnarray}
where the cavity field and mechanical oscillator modes are described by their effective frequencies, $\omega_{c}$ and $\omega_{m}$, and creation (annihilation) operators, $\hat{a}^{\dagger}$ ($\hat{a}$ )   and $\hat{b}^{\dagger}$ ($\hat{b}$ ), in that order.
Dispersive linear coupling between the modes occurs and it is quantified by the coupling constant $g$.
Here, we will revisit this and aggregated models through an operator approach.
In the following section, we will show how the well know equivalence between the standard optomechanical model and a Kerr medium. 
We will also show that the driven optomechanical model is similar to a trapped ion and that is the reason behind the use of sideband cooling and other ion-trap cavity electrodynamics (QED) techniques to prepare mechanical states.
We will couple the mechanical mode to an environment and introduce super-operator techniques to make the problem tractable. 
We will introduce a new result showing that the open system reduces to a damped mechanical oscillator  in the case of a thermal electtromagnetic field mode in the cavity.
Finally, we will add a two-level system, interacting with the cavity field under Jaynes-Cummings dynamics, to the model and show that a right unitary approach allows us to understand how the electromagnetic field mode mediates coupling between the atom and the mechanical oscillator but a closed form time evolution operator requires developing an adequate rotating wave approximation (RWA) compatible with the right unitary transformations.

\section{Standard optomechanical model}

The first experimental realization of a classical optomechanical cavity consisted of a Fabry-Perot cavity with a fixed mirror and a pendulum-like moving mirror \cite{Dorsel1983p1550}. 
In this experiment and other proposals \cite{Meystre1985p1830}, optical bistability and mirror confinement due to changes in the physical length of the cavity induced by radiation pressure were shown. 
This bistable phonemenon was similar to that found in fixed cavities filled with nonlinear media \cite{Marburger1978p335}.
Soon, it was shown that the equations of motion of the quantum optomechanical system showed optical bistability due to its equivalence to a Kerr medium \cite{Hilico1992p202,Jackel1992p39}.
The topic has been revisited through different approaches after that \cite{RodriguezLara2004p354,Aldana2013p043826}.

\subsection{Driven system}
Here we are interested in an algebraic approach, for this reason we will start with the standard Hamiltonian for a pumped optomechanical system,
\begin{equation}\label{hcm}
\hat{H}_{p}=\omega_{c}\hat{a}^{\dag}\hat{a}+\omega_{m}\hat{b}^{\dag}\hat{b}-g\hat{a}^{\dag}\hat{a}\left(\hat{b}^{\dag}+\hat{b}\right) + \Omega \cos\left(\omega_{p}t\right) \left(\hat{a}^{\dag}+\hat{a}\right),
\end{equation}
where the laser pump frequency is given by $\omega_{p}$ and its strength by $\Omega$.
First, we need to get rid of the time dependence, so we move to a frame defined by the cavity field photon number rotating at the pump frequency, $\vert \psi \rangle = \hat{U}_{R}(t) \vert \phi \rangle$ with  $\hat{U}_{R}(t) = e^{- i \omega_{p} \hat{a}^{\dagger} \hat{a} t}$, such that we can write the Schr\"odinger equation as, 
\begin{eqnarray}
i \partial_{t} \left[ \hat{U}_{R}(t) \vert \phi \rangle \right] = \hat{H}_{p} \hat{U}_{R}(t) \vert \phi \rangle.
\end{eqnarray}
Thus the effective Hamiltonian in the new frame,
\begin{eqnarray}
i \partial_{t}  \vert \phi \rangle =  \hat{U}_{R}^{\dagger}(t) \left[ \hat{H}_{p} -  \omega_{p} \hat{a}^{\dagger} \hat{a} \right] \hat{U}_{R}(t) \vert \phi \rangle,
\end{eqnarray}
 is given by the following,
\begin{eqnarray}
\fl \hat{H}_{R} &=& \hat{U}_{R}^{\dagger}(t) \left[ \hat{H}_{p} -  \omega_{p} \hat{a}^{\dagger} \hat{a} \right] \hat{U}_{R}(t), \nonumber \\
\fl &=&  \delta_{p} \hat{a}^{\dag}\hat{a} + \omega_{m} \hat{b}^{\dag} \hat{b}- g \hat{a}^{\dag} \hat{a} \left(\hat{b}^{\dag}+\hat{b}\right) + \frac{\Omega}{2}  \left[ \hat{a}^{\dag} (1 + e^{2 i \omega_{p} t} ) + \hat{a} (1 + e^{-2 i \omega_{p} t} )\right]. 
\end{eqnarray}
Finally, we can make a RWA to eliminate the terms rotating at twice the pump frequency, $e^{\pm 2 i \omega_{p} t}$, and obtain, 
\begin{eqnarray}
\hat{H}_{c} = \delta_{p}\hat{a}^{\dag}\hat{a}+\omega_{m}\hat{b}^{\dag}\hat{b}-g\hat{a}^{\dag}\hat{a}\left(\hat{b}^{\dag}+\hat{b}\right)+\frac{\Omega}{2}\left(\hat{a}^{\dag}+\hat{a}\right),
\end{eqnarray}
where the detuning between the pump and cavity field frequencies is $\delta_{p} = \omega_{c} - \omega_{p}.$
Now, we want to get rid of the terms involving the cavity field intensity and the canonical position of the mechanical oscillator.
For this, let us define a displacement on the mechanical oscillator basis, 
\begin{eqnarray}
\hat{D}_{\hat{b}}( \hat{\xi} )= e^{ \left(\hat{\xi}\hat{b}^{\dag} - \hat{\xi}^{\dagger}\hat{b}\right)},
\end{eqnarray}
where the operator $\hat{\xi}$ is either a cavity field operator or just a complex number.
If we change into a joint basis defined by such a displacement, we arrive to a Hamiltonian closer to our goal \cite{Bose1997p4175,Ludwig2012p063601},
\begin{eqnarray}
\hat{H}_{D} & = & \hat{D}_{\hat{b}}^{\dagger} \left( \frac{g}{\omega_{m}} \hat{a}^{\dagger} \hat{a} \right)\hat{H}_{c}\hat{D}_{\hat{b}} \left(  \frac{g}{\omega_{m}} \hat{a}^{\dagger} \hat{a} \right), \nonumber \\
&=& \delta_{p}\hat{a}^{\dag}\hat{a}  - \frac{g^{2}}{\omega_{m}} \left(\hat{a}^{\dag}\hat{a}\right)^{2} + \omega_{m}\hat{b}^{\dag}\hat{b} + \frac{\Omega}{2} \left[ \hat{a}^{\dag} \hat{D}_{\hat{b}}^{\dagger} \left(  \frac{g}{\omega_{m}} \right) + \hat{a}  \hat{D}_{\hat{b}}\left(  \frac{g}{\omega_{m}} \right) \right].
\end{eqnarray}
Note that moving into the joint displaced basis defined by the operator $\hat{D}( g \hat{a}^{\dagger} \hat{a} / \omega_{m})$ helps us to get rid of the linear dispersive mechanical-cavity field modes interaction but introduces a Kerr term for the cavity field, $\left( \hat{a}^{\dagger} \hat{a} \right)^{2}$, and switches the interaction to the driving terms that become $\hat{a}^{\dagger} \hat{D}^{\dagger}_{\hat{b}}(g/\omega_{m})$ and $\hat{a} \hat{D}_{\hat{b}}(g/\omega_{m})$.
In order to reach our goal, we now move to the rotating frames defined by the intensity in the cavity field rotating at the detuning frequency and the number of excitations in the mechanical mode rotating with frequency $\omega_{m}$, $ \hat{U}_{cm}(t)=e^{\left[-i\left(\delta_{p}\hat{a}^{\dag}\hat{a}+\omega_{m}\hat{b}^{\dag}\hat{b}\right)t\right]}$.
Thus, we reach our goal, 
\begin{eqnarray}
\fl  \hat{H}_{cm} &=&  \hat{U}_{cm}^{\dagger}(t) \left(\hat{H}_{D}-\delta_{p}\hat{a}^{\dag}\hat{a}-\omega_{m}\hat{b}^{\dag}\hat{b}\right)\hat{U}_{cm}(t)  \nonumber \\
\fl  &=& -\frac{g^{2}}{\omega_{m}}\left(\hat{a}^{\dag}\hat{a}\right)^{2}+\frac{\Omega}{2}\left[ \hat{a}^{\dag} e^{i\delta_{p}t} \hat{D}^{\dagger}_{\hat{b}}\left(\frac{g}{\omega_{m}} e^{i\omega_{m}t}\right) + \hat{a} e^{-i\delta_{p}t} \hat{D}_{\hat{b}}\left(\frac{g}{\omega_{m}} e^{i\omega_{m}t}\right) \right]. 
\end{eqnarray}
Note that in the absence of driving, $\Omega = 0$, the standard optomechanical Hamiltonian is equivalent to a Kerr Hamiltonian \cite{Hilico1992p202,Jackel1992p39,RodriguezLara2004p354,Aldana2013p043826} and it has been shown that measuring the field quadratures of this system delivers information about the Wigner characteristic function of the mechanical oscillator \cite{RodriguezLara2004p213}.
Furthermore, from such a form it is straightforward to discuss light squeezing \cite{SafaviNaeini2013p185}, photon blockade \cite{Rabl2011p063601} and single photon dynamics \cite{Tang2014p063821}, to mention a few examples.

In the presence of driving, the second term in the effective Hamiltonian $\hat{H}_{cm}$ is quite interesting.
Note how similar these terms are to that of a driven trapped ion \cite{MoyaCessa2012p229}; as a matter of fact, if we substituted the cavity field operators by Pauli matrices we would recover the trapped ion Hamiltonian.
Thus, we can use an approach similar to that used in trapped-ion QED and expand the mechanical mode operators into their power series, 
\begin{eqnarray}
\hat{D}^{\dagger}_{\hat{b}}\left( \alpha e^{i\omega_{m}t} \right) &=& e^{- \frac{\vert \alpha \vert^{2}}{2}} \sum_{p,q=0}^{\infty} \frac{1}{p!q!} (-\alpha \hat{b}^{\dagger})^{p} (\alpha \hat{b})^{q} e^{i \omega_{m} \left( q - p \right) t}, \quad \alpha \equiv \frac{g}{\omega_{m}}, \\
\hat{D}_{\hat{b}}\left( \alpha e^{i\omega_{m}t} \right) &=& e^{- \frac{\vert \alpha \vert^{2}}{2}} \sum_{p,q=0}^{\infty} \frac{1}{p!q!} (\alpha \hat{b}^{\dagger})^{p} (-\alpha \hat{b})^{q} e^{-i \omega_{m} \left( q - p \right) t}.
\end{eqnarray}
If the displacement parameter, $\alpha = g/ \omega_{m}$, fulfills $\alpha \ll 1$, the pump intensity is high, $\Omega \gg \omega_{c}$, and we choose the pumping detuning such that it is an integer multiple of the mirror frequency, $\delta_{p} = \pm s \omega_{m}$ with $s=0,1,2,\ldots$, then we can apply a RWA to obtain, for $\delta = s \omega_{m}$,
\begin{eqnarray}
\hat{H}_{+} &=& -\frac{g^{2}}{\omega_{m}}\left(\hat{a}^{\dag}\hat{a}\right)^{2}+ \frac{\Omega}{2} e^{- \frac{\vert \alpha \vert^{2}}{2}} \left[ \hat{a} \frac{(\hat{b}^{\dagger}\hat{b})!}{(\hat{b}^{\dagger} \hat{b} +s)!} L^{(s)}_{\hat{b}^{\dagger}\hat{b}}( \alpha^{2})(\alpha \hat{b})^{s} + \right. \nonumber \\
&& \left.  + \hat{a}^{\dagger}(-\alpha \hat{b}^{\dagger})^{s} \frac{(\hat{b}^{\dagger} \hat{b})!}{(\hat{b}^{\dagger}\hat{b}+s)!} L^{(s)}_{\hat{b}^{\dagger}\hat{b}}( \alpha^{2}) \right],
\end{eqnarray}
and for $\delta = - s \omega_{m}$,
\begin{eqnarray}
\hat{H}_{-} &=& -\frac{g^{2}}{\omega_{m}}\left(\hat{a}^{\dag}\hat{a}\right)^{2}+ \frac{\Omega}{2} e^{- \frac{\vert \alpha \vert^{2}}{2}} \left[ \hat{a} (-\alpha \hat{b}^{\dagger})^{s} \frac{(\hat{b}^{\dagger}\hat{b})!}{(\hat{b}^{\dagger}\hat{b}+s)!} L^{(s)}_{\hat{b}^{\dagger}\hat{b}}( \alpha^{2}) + \right. \nonumber \\
&& \left.  + \hat{a}^{\dag}  \frac{(\hat{b}^{\dagger}\hat{b})!}{(\hat{b}^{\dagger}\hat{b}+s)!} L^{(s)}_{\hat{b}^{\dagger}\hat{b}}(\alpha^{2})(\alpha \hat{b})^{s} \right],
\end{eqnarray}
where the function $L^{(m)}_{n}(x)$ is the $n$-th Laguerre generalized polynomial with parameter $m$.
In other words, by choosing the detuning between the cavity and pump fields we can produce a nonlinear coupling of the cavity field with the mechanical oscillator in a form equivalent to that of a trapped-ion.
Thus, we can use the knowledge from ion-trap QED to engineer quantum states in the closed or open standard optomechanical system \cite{Mancini2002p120401,Bhattacharya2008p030303,Marshall2003p130401,Hong2013p023812,Akram2013p093007,Xu2013p053849,Xu2013p063819}; e.g. Sideband cooling \cite{Bhattacharya2007p073601,Marquardt2007p093902,WilsonRae2007p093901}, Schr\"odinger cats \cite{Bose1997p4175,Mancini1997p3042}, non-Gaussian states of the mechanical oscillator \cite{Khalili2010p070403,Gu2013p01385,Kronwald2013p133601,Gu2014p18254}.

\subsection{Damping of the mechanical oscillator}

Now we will turn our attention to the standard optomechanical system and take into account the damping of the mechanical oscillator, such that the system dynamics is described by a master equation,
\begin{eqnarray}
\frac{d}{dt} \hat{\rho} = - i \left[ \hat{H} , \hat{\rho} \right] + \gamma \hat{\mathcal{L}}_{\hat{b}} \left[ \hat{\rho} \right],
\end{eqnarray} 
where $\gamma$ is the decay rate and $\hat{\rho}$ is the density operator of the system and the Linblad superoperator is given by, 
\begin{eqnarray}
\hat{\mathcal{L}}_{\hat{\zeta}}\left[ \hat{\rho} \right] = 2 \hat{\zeta} \hat{\rho} \hat{\zeta}^{\dagger} - \hat{\zeta}^{\dagger} \hat{\zeta} \hat{\rho} - \hat{\rho} \hat{\zeta}^{\dagger} \hat{\zeta}.
\end{eqnarray}
It is straightforward to move into the frame defined by the photon number rotating at the cavity field frequency, $\hat{U}_{c}(t) = e^{-i \omega_{f} \hat{a}^{\dagger} \hat{a} t }$, then, the master equation becomes 
\begin{eqnarray}
\frac{d}{dt} \hat{\rho}_{c} = - i \left[ \hat{H}_{m} , \hat{\rho}_{c} \right] + \gamma \hat{\mathcal{L}}_{\hat{b}} \left[ \hat{\rho}_{c} \right], \quad \hat{H}_{m} = \omega_{m} \hat{b}^{\dagger} \hat{b} + g \hat{a}^{\dagger} \hat{a}  \left( \hat{b}^{\dagger} + \hat{b} \right),
\end{eqnarray} 
and $\hat{\rho}_{c} = \hat{U}_{c}(t) \hat{\rho} \hat{U}_{c}^{\dagger}(t)$.
Now, we follow the standard procedure mentioned above and introduce a displacement of the mechanical degree of freedom as a function of the number of photons in the cavity, $\hat{D} \left(\beta \hat{a}^{\dagger} \hat{a} \right)$ with $\beta \in \mathbb{C}$.
Thus, we can write the density operator as
\begin{eqnarray}
\hat{\rho} = \hat{D} \left(\beta \hat{a}^{\dagger} \hat{a}  \right) \hat{\rho}_{D} \hat{D}^{\dagger} \left(\beta \hat{a}^{\dagger} \hat{a}  \right),
\end{eqnarray}
and its corresponding master equation, 
\begin{eqnarray}
\fl \frac{d}{dt} \hat{\rho}_{D} &=& - i \left[ \hat{H}_{D} , \hat{\rho}_{D} \right] + \gamma \left\{ \hat{\mathcal{L}}_{\hat{b}} \left[ \hat{\rho}_{D} \right] +  \hat{\mathcal{L}}_{\hat{a}^{\dagger} \hat{a} } \left[ \hat{\rho}_{D} \right] +  \beta \hat{\mathcal{N}}\left[\hat{\rho}_{D}\right] + \beta^{\ast} \hat{\mathcal{N}}^{\dagger}\left[\hat{\rho}_{D}\right]   \right\} ,
\end{eqnarray}
where the displaced Hamiltonian is
\begin{eqnarray}
\hat{H}_{D} =  \epsilon \left( \hat{a}^{\dagger} \hat{a} \right) ^{2} + \omega_{m} \hat{b}^{\dagger} \hat{b} + \hat{a}^{\dagger} \hat{a}  \left( \mu \hat{b}^{\dagger} + \mu^{\ast} \hat{b} \right)
\end{eqnarray}
with parameters, 
\begin{eqnarray}
\mu &=& g + \omega \beta, \\
\epsilon &=& 2 g \mathrm{Re}(\beta) + \omega \vert \beta \vert^{2},
\end{eqnarray}
and the new superoperator is defined in the following,
\begin{eqnarray}
\hat{\mathcal{N}}\left[\hat{\rho}\right] = \left( 2 \hat{a}^{\dagger} \hat{a}  \hat{\rho} \hat{b}^{\dagger} - \hat{b}^{\dagger} \hat{a}^{\dagger} \hat{a}  \hat{\rho} - \hat{\rho} \hat{b}^{\dagger} \hat{a}^{\dagger} \hat{a} \right).
\end{eqnarray}

At this point, we can start analyzing which initial conditions of the field provides us with a master equation amenable for analytic solution.
The simplest case is given by a thermal field in the cavity with an average of $\bar{n}$ photons, 
\begin{eqnarray}
\hat{\rho}_{th}(\bar{n}) = \sum_{k=0}^{\infty} \frac{\bar{n}^{k}}{(1 + \bar{n})^{k+1}} \vert k \rangle \langle k \vert,
\end{eqnarray}
leading to the following initial state of the whole system,
\begin{eqnarray}
\hat{\rho}_{D}(0) = \hat{\rho}_{th}(\bar{n}) \otimes \hat{\rho}_{m}(0).
\end{eqnarray}
Then, we can set the value of the displacement parameter, 
\begin{eqnarray}
\beta = - \frac{g}{ \omega_{m} - i \gamma},
\end{eqnarray}
to get the following master equation for this particular case of initial density operator, 
\begin{eqnarray}
\frac{d}{dt} \hat{\rho}_{D} &=& - i \left[ \omega_{m} \hat{b}^{\dagger} \hat{b}, \hat{\rho}_{D} \right] + \gamma \mathcal{L}_{\hat{b}} \left[\hat{\rho}_{D}\right].
\end{eqnarray}
In other words, a mechanical oscillator interacting with a thermal field cavity mode and coupled to an environment behaves just as a free mechanical harmonic oscillator coupled to an environment.
Its time evolution can be given by standard superoperator techniques \cite{Phoenix1990p5132,ArevaloAguilar1996p675,ArevaloAguilar1998p671,Lu2003p024101},
\begin{equation}
\hat{\rho}_D(t)= e^{\hat{L}t} e^{\frac{1-e^{-2\gamma }}{2\gamma}\hat{J}t} \left[ e^{-i \omega_{m} \hat{b}^{\dagger}\hat{b}  t}\hat{\rho}_D(0) e^{i \omega_{m} \hat{b}^{\dagger} \hat{b}  t}\right],
\end{equation}
with the auxiliary superoperators,
\begin{equation}
\hat{J}\hat{\rho}=2\gamma \hat{b} \hat{\rho} \hat{b}^{\dagger}, \qquad \hat{L}\hat{\rho}=-\gamma  (b^{\dagger}b\hat{\rho} +\hat{\rho}b^{\dagger}b).
\end{equation}


\section{Hybrid qubit-optomechanical model}

Recently, it has been proposed to couple a two-level atom to the standard optomechanical model \cite{Ian2008p013824,Genes2008p050307,Pflanzer2013p033804}, such an hybrid model is described by the Hamiltonian, 
\begin{eqnarray}
\hat{H}_{h} = \omega_{c}\hat{a}^{\dag}\hat{a}+\omega_{m}\hat{b}^{\dag}\hat{b}-g\hat{a}^{\dag}\hat{a}\left(\hat{b}^{\dag}+\hat{b}\right) + \frac{\omega_{0}}{2} \hat{\sigma}_{z} + \lambda \left( \hat{a}^{\dagger} \hat{\sigma}_{+} + \hat{a} \hat{\sigma}_{-} \right),
\end{eqnarray}
where the two-level system is described by Pauli matrices, $\hat{\sigma}_{j}$ with $j=z,\pm$, and the transition frequency $\omega_{0}$, and the atom-field coupling is given by the parameter $\lambda$.
Here, we will first move into the frame defined by the photon number and the qubit energy rotating at the cavity field frequency, $\hat{U}_{r} = e^{ -i \omega_{c} \left( \hat{a}^{\dagger} \hat{a} + \hat{\sigma}_{z} /2 \right)t}$, such that the effective Hamiltonian is, 
\begin{eqnarray}
\hat{H}_{r} = \frac{\delta}{2} \hat{\sigma}_{z} + \omega_{m} \hat{b}^{\dagger} \hat{b} + \lambda \left( \hat{a}^{\dagger} \hat{\sigma}_{+} + \hat{a} \hat{\sigma}_{-} \right) - g \hat{a}^{\dagger} \hat{a} \left(\hat{b}^{\dagger} + \hat{b} \right),
\end{eqnarray}
where the detuning between the qubit and cavity field frequency is given by $\delta = \omega_{0} - \omega_{c}$.
We can follow a right unitary approach \cite{Tang1996p154,RodriguezLara2011p3770,RodriguezLara2013p095301} and rewrite this Hamiltonian in the form, 
\begin{eqnarray}
\hat{H}_{r} = \hat{T}\hat{H}_{T}\hat{T}^{\dag},
\end{eqnarray}
where the auxiliary Hamiltonian is given by the expression,
\begin{eqnarray}
\hat{H}_{T} = \frac{\delta}{2}\hat{\sigma}_{z}+ \omega_{m}\hat{b}^{\dag}\hat{b} + \lambda\sqrt{\hat{a}^{\dag}\hat{a}}~\hat{\sigma}_{x}-g\hat{a}^{\dag}\hat{a}\left(\hat{b}^{\dag}+\hat{b}\right) + g\left(\hat{b}^{\dag}+\hat{b}\right)\vert e \rangle \langle e\vert, \label{eq:HamT}
\end{eqnarray}
where we have diagonalized the cavity field part by using the operators, 
\begin{equation}
\hat{T}= \left( \begin{array}{ccc} \hat{V} & 0\\ 0 & 1 \end{array} \right), \quad
\hat{T}^{\dag}=\left( \begin{array}{ccc} \hat{V}^{\dag} & 0 \\ 0 & 1 \end{array}  \right),
\end{equation}
are right unitary due to the properties of the Susskind-Glogower operators, 
\begin{eqnarray}
\hat{V}=\frac{1}{\sqrt{\hat{a}^{\dag}\hat{a}+1}} \hat{a}, \quad
\hat{V}^{\dag}= \hat{a}^{\dag} \frac{1}{\sqrt{\hat{a}^{\dag}\hat{a}+1}},
\end{eqnarray} 
that fulfill $\hat{V} \hat{V}^{\dagger} = 1$ but $\hat{V}^{\dagger} \hat{V} = 1 - \vert 0 \rangle \langle 0 \vert$.
Note that we can re-arrange the following terms in the auxiliary Hamiltonian,
\begin{equation}
\fl g \left(\hat{b}^{\dag}+\hat{b}\right) \vert e \rangle \langle e \vert - g \hat{a}^{\dag} \hat{a} \left( \hat{b}^{\dag} + \hat{b}\right) =\frac{1}{2}g\left(\hat{b}^{\dag}+\hat{b}\right)\hat{\sigma}_{z}-g\left(\hat{a}^{\dag}\hat{a}-\frac{1}{2}\right)\left(\hat{b}^{\dag}+\hat{b}\right),
\end{equation}
and use again the displacement operator in terms of the number of photon in the field to obtain,
\begin{eqnarray}
\hat{H}_{d} &=& \hat{D}^{\dagger} \left[ \frac{g}{\omega_{m}} \left( \hat{a}^{\dagger} \hat{a} - \frac{1}{2} \right)\right] \hat{H}_{T} \hat{D}\left[ \frac{g}{\omega_{m}} \left( \hat{a}^{\dagger} \hat{a} - \frac{1}{2} \right)\right], \\
 &=& \left[ \frac{\delta}{2} + \frac{g}{2} \left(\hat{b}^{\dagger}+\hat{b}\right) + \frac{g^{2}} {\omega_{m}} \left(\hat{a}^{\dagger} \hat{a} - \frac{1}{2} \right) \right] \hat{\sigma}_{z} + \omega_{m}\hat{b}^{\dagger} \hat{b}  \nonumber \\
 && + \lambda\sqrt{\hat{a}^{\dagger} \hat{a}} \hat{\sigma}_{x} - \frac{g^{2}}{\omega_{m}} \left(\hat{a}^{\dagger}\hat{a} - \frac{1}{2}\right)^{2}.
\end{eqnarray}
Now, just for the sake of clarity, we can introduce a rotation around $\hat{\sigma}_{y}$, 
\begin{eqnarray}
\hat{R}_{y}\left( \theta \right) = e^{ - i \theta \hat{\sigma}_{y} },
\end{eqnarray}
and rewrite our initial hybrid optomechanical Hamiltonian as,
\begin{eqnarray}
\fl H_{r} = \hat{T}  \hat{D}_{\hat{b}} \left[ \frac{g}{\omega_{m}} \left( \hat{a}^{\dagger} \hat{a} - \frac{1}{2} \right)\right] \hat{R}_{y}^{\dagger}\left( \frac{\pi}{4} \right) \hat{H}_{a} \hat{R}_{y}\left( \frac{\pi}{4} \right) \hat{D}_{\hat{b}}^{\dagger}\left[ \frac{g}{\omega_{m}} \left( \hat{a}^{\dagger} \hat{a} - \frac{1}{2} \right)\right] \hat{T}^{\dagger}, 
\end{eqnarray}
with the final auxiliary Hamiltonian given by,
\begin{eqnarray}
\hat{H}_{a} &=& \hat{H}_{K}  + \hat{H}_{am}.
\end{eqnarray}
Note that the effective  Kerr medium, 
\begin{eqnarray}
\hat{H}_{K}= - \frac{g^{2}}{\omega_{m}} \left( \hat{a}^{\dagger} \hat{a} - \frac{1}{2}\right)^{2},
\end{eqnarray}
commutes with the rest of the terms, 
\begin{eqnarray}
\hat{H}_{am} &=& \omega_{m} \hat{b}^{\dagger} \hat{b}  + \tilde{\omega}\left( \hat{a}^{\dagger} \hat{a} \right) \hat{\sigma}_{z} + \tilde{\Omega}\left( \hat{a}^{\dagger} \hat{a} \right) \hat{\sigma}_{x}  + \frac{g}{2} \left(\hat{b}^{\dagger} + \hat{b} \right) \hat{\sigma}_{x},
\end{eqnarray}  
that can reinterpreted as a driven two-level atom interacting with the mechanical oscillator under the Jaynes-Cummings model \cite{Jaynes1963p89} without the RWA. 
The two-level transition frequency and driving strength depend on the intensity of the optical mode,
\begin{eqnarray}
\tilde{\omega}\left( \hat{a}^{\dagger} \hat{a} \right) &=& -  \lambda \sqrt{\hat{a}^{\dagger} \hat{a}}, \\
\tilde{\Omega}\left( \hat{a}^{\dagger} \hat{a} \right) &=&  \frac{\delta}{2} + \frac{g^{2}}{2 \omega_{m}} \left( \hat{a}^{\dagger} \hat{a} - \frac{1}{2}\right) .
\end{eqnarray}

At this point we could note that, 
\begin{eqnarray}
\fl \hat{H}_{r}^{2} &=& \hat{T} \hat{H}_{T}^{2} \hat{T}^{\dagger} - \hat{T} \hat{H}_{T} \left( \vert e \rangle \langle e \vert \otimes \vert 0 \rangle \langle 0 \vert \otimes \sum_{k=0}^{\infty} \vert k \rangle \langle k \vert \right) \hat{H}_{T} \hat{T}^{\dagger}, \nonumber \\
\fl  &=& \hat{T} \hat{H}_{T}^{2} \hat{T}^{\dagger} - \left[\frac{\delta}{2} + \omega_{m} k + g \left( \hat{b} + \hat{b}^{\dagger} \right) \right] \left( \vert e \rangle \langle e \vert \otimes \hat{V} \vert 0 \rangle \langle 0 \vert \hat{V}^{\dagger} \otimes \sum_{k=0}^{\infty} \vert k \rangle \langle k \vert \right)  \nonumber \\
\fl  && \left[\frac{\delta}{2} + \omega_{m} k + g \left( \hat{b} + \hat{b}^{\dagger} \right) \right], \nonumber \\
\fl  &=& \hat{T} \hat{H}_{T}^{2} \hat{T}^{\dagger},
\end{eqnarray}
and write the evolution operator of the total system as, 
\begin{eqnarray}
\fl \hat{U}(t) = \hat{T}  \hat{D}_{\hat{b}} \left[ \frac{g}{\omega_{m}} \left( \hat{a}^{\dagger} \hat{a} - \frac{1}{2} \right)\right] \hat{R}_{y}\left( \frac{\pi}{4} \right) e^{-i \hat{H}_{a} t } \hat{R}_{y}^{\dagger}\left( \frac{\pi}{4} \right) \hat{D}_{\hat{b}}^{\dagger}\left[ \frac{g}{\omega_{m}} \left( \hat{a}^{\dagger} \hat{a} - \frac{1}{2} \right)\right] \hat{T}^{\dagger}, 
\end{eqnarray}
where we can use the fact that the Kerr term commutes, 
\begin{eqnarray}
e^{-i \hat{H}_{a} t } = e^{-i \hat{H}_{K} t } e^{-i \hat{H}_{am} t },
\end{eqnarray}
but in the end, it is not possible to provide a closed form propagator from the term $e^{-i \hat{H}_{am} t }$ and working a formal rotating wave approximation in this scenario is beyond our current purpose.

\section{Conclusion}

We used a purely algebraic approach to revisit the standard quantum optomechanical model describing the linear dispersive interaction between two bosonic modes, electromagnetic and mechanical. 
We showed that a displacement on the mechanical basis proportional to the number state in the electromagnetic basis, sometimes called a polaron transformation, delivers a model consisting of an effective electromagnetic Kerr medium plus a coupling between the electromagnetic and mechanical modes similar to that found in a trapped ion. 
We took advantage of ion-trap-QED and showed that it is possible to implement a series of optical-mechanical couplings that allow trapping, cooling and parametric coupling phenomena, by choosing detunings between the electromagnetic mode and the classical pump that are integer multiples of the mechanical frequency.

We also used super-operator techniques to revisit the standard optomechanical system when the mechanical oscillator is coupled to the environment. 
Here we worked out a general expression and gave a closed form time evolution super-operator for the particular case of a thermal electromagnetic field. 

Finally, we presented a right unitary approach to a system formed by the addition of a two-level atom interacting with the electromagnetic mode. 
This approach makes it simple to realize that the electromagnetic mode enables the coupling between the two-level system with the mechanical mode in a Jaynes-Cummings without the RWA form but also shows us that it is not possible to provide a closed form time evolution operator unless an adequate approximation scheme is developed.

%
\section*{Acknowledgments}
C. Ventura Vel\'azquez acknowledges financial support from CONACyT through the master degree scholarship $\#$294810.

%
\section*{References}

\providecommand{\newblock}{}

\end{document}